\documentclass[jkps,preprint,fleqn,showpacs,showkeys]{revtex4}
\usepackage{graphicx}
\usepackage{amssymb}
\usepackage{amsmath}
\usepackage{bm}

\begin{document}

\title[]{Can the Horowitz-Maldacena Proposal be an\\ Alternative to the Firewall?}
\author{Dong-han \surname{Yeom}}
\email{innocent.yeom@gmail.com}
\affiliation{\small{Center for Quantum Spacetime,\\ Sogang University, Seoul 121-742, Korea,}}
\affiliation{\small{Yukawa Institute for Theoretical Physics, Kyoto University, Kyoto 606-8502, Japan}}
\author{Heeseung \surname{Zoe}}
\affiliation{\small{Division of General Studies,\\ Ulsan National Institute of Science and Technology,\\ Ulsan 689-798, Korea}}

\begin{abstract}
Recently, there have been discussions that black hole complementarity is inconsistent and that the firewall is required to prohibit the observation of duplicated information. It is thought that if the Horowitz-Maldacena proposal works as a selection principle, then this may be an alternative to the firewall. In this paper, we first point out that the Horowitz-Maldacena proposal seems to help black hole complementarity for charged black holes. However, if we consider the Hayden-Preskill argument further, which states that a black hole can function as an information mirror after the information retention time, then we can show that the Horowitz-Maldaceana proposal cannot help black hole complementarity. This can be extended to neutral black hole cases. Therefore, in conclusion, we find that dynamical black holes do not respect complementarity, even with the Horowitz-Maldacena proposal.
\end{abstract}

\preprint{YITP-13-71}



\maketitle

{\centering
\section{\label{sec:intro}Introduction}
}

Many ideas for resolving the black hole information loss paradox have been suggested \cite{inforpara}.
One of the most important proposals is known as the \textit{black hole complementarity principle}, whereby asymptotic and free-falling observers \textit{see} different things while known physical laws, especially unitarity, remain preserved \cite{complementarity}.
However, quantum information that consists of black hole internal states to be observed by the free-falling observer seems to be copied to outgoing Hawking radiation directed towards the asymptotic observer.
This duplication of quantum information violates the \textit{no quantum Xeroxing} theorem.
The black hole complementarity principle argues that the viewpoints of the two observers are essentially complementary.
This poses no problem (in terms of the above information duplication) because the observers can never communicate.
It is known that, if we assume the statistical entropy to be equal to thermal entropy \cite{Strominger:1996sh} while keeping the unitarity of quantum mechanics intact \cite{Maldacena:1997re}, then the black hole complementarity principle would be manifest \cite{nonlocal} due to quantum information theoretic arguments \cite{Page}.

However, one needs to further check the consistency of black hole complementarity;
\textit{is it really impossible to observe the duplication of information?}
This impossibility was confirmed only for the \textit{specific} case of a Schwarzschild black hole with limited assumptions \cite{inforretention}.
However, the authors did argue that complementarity might not hold if local horizons become different from global horizons.
Numerical calculations of charged black holes indicate a distance separation between the inner and the Cauchy horizons \cite{HHSY}.
This intervening space allows for gedanken experiments that can probe whether or not information duplication is possible \cite{HHYZ}.
To work as a good counterexample, in Refs. \cite{HHSY,Yeom:2008qw}, a large number of scalar fields was assumed. On the other hand, this assumption could be reduced to a reasonable number that could be justified in string theory \cite{Yeom:2009zp,Kim:2013fv}. Furthermore, recently, the inconsistency of black hole complementarity was observed by other authors, and the firewall was introduced to prevent duplication of information \cite{Almheiri:2012rt,Hwang:2012nn}.

Fortunately, however, a loophole existed, for it had been suggested that dynamical black holes might violate black hole complementarity through duplication experiments only when no relation existed between the singularity and the outer horizon.
If a special \textit{conspiracy} between the singularity and the outer horizon could be implemented via quantum teleportation,
a dynamical black hole might rescue black hole complementarity.
Here, the Horowitz-Maldacena proposal \cite{HM} could function as a selection principle by using the quantum teleportation \cite{HHYZ}. Then, this could be an alternative of the firewall.

In this paper, we argue that this resolution is \textit{insufficient} in terms of rescuing complementarity (i.e., we will re-consider a resolution of Ref. \cite{HHYZ}). Because it is more useful to intuitively understand, in Sec.~\ref{sec:dup}, we summarize the duplication experiment for a charged black hole. In Sec.~\ref{sec:HM}, we suggest that the Horowitz-Maldacena proposal can partly rescue the principle of black hole complementarity. However, in Sec.~\ref{sec:HP}, by consideration of the Hayden-Preskill argument \cite{Hayden:2007cs}, we show that we can construct a situation whereby the selection principle still allows for duplication experiments. In Sec.~\ref{sec:gen}, we generalize to a neutral black hole to reduce the assumption of a large number of scalar fields. In Sec.~\ref{sec:dis}, we summarize our results. Thus, we have justification for our claim that dynamical charged/neutral black holes unambiguously violate black hole complementarity.

\begin{figure}
\begin{center}
\includegraphics[scale=0.7]{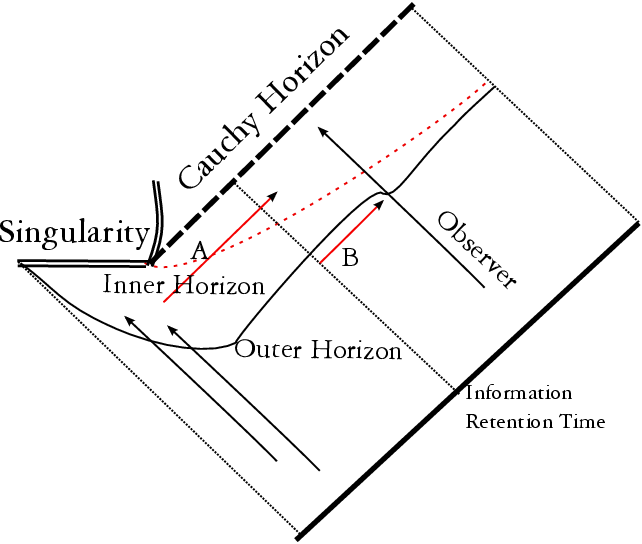}
\caption{\label{fig1}
This figure can be obtained by pasting the mass inflation scenario to the extreme black hole.
One can see the space-like bending of the inner horizon, as well as the Cauchy horizon, starting from the end of the space-like singularity.
If the final observer can see both $A$ and $B$, then the observer may see a duplication of information.}
\end{center}
\end{figure}

~\\

{\centering
\section{\label{sec:dup}Duplication experiment in a charged black hole}
}

If we assume suppression of pair-creation, the causal structure of a dynamical charged black hole can be obtained by pasting from the mass inflation scenario \cite{Poisson:1990eh} to the extreme black hole solution.
In the mass inflation scenario, the outer apparent horizon grows along the space-like direction, and a space-like singularity exists due to mass inflation \cite{Bonanno:1994ma}.
If no Hawking radiation exists, the inner horizon sits at an infinite advanced time $v \rightarrow \infty$ \cite{Ori,HodPiran}.
The mass function behaves as $m(u,v) \sim \exp{\kappa_{i} (u+v)}$, where $\kappa_{i}$ is the surface gravity of the inner horizon,
and $u$ and $v$ are the retarded and the advanced time parameters for the double null coordinate \cite{Poisson:1990eh}. Thus, the inner horizon becomes a curvature singularity.
However, if we paste this scenario to the extreme black hole, the inner horizon must approach the outer horizon.
Then, the most natural guess is that the inner horizon bends in a space-like direction and approaches the outer horizon;
this expectation has been confirmed by the authors through numerical calculations \cite{SorkinPiran,HHSY}.
Notice that one can access any location in the integrated domain by using finite $u$ and $v$.
This implies that no curvature singularity exists in the general relativistic sense.
Of course, because the mass function diverges exponentially, some curvature functions diverge on a scale than the Planck scale.
This is resolved by re-scaling the unit length (i.e., Planck length), and this re-scaling can be implemented by tuning the number of massless degrees of freedom \cite{HHSY}.
Finally, if we assume a large number of massless degrees of freedom, we can get the semiclassically convincing causal structure of a dynamical charged black hole (Fig.~\ref{fig1}).

One important point is that the transition region between the mass inflation scenario and the extreme black hole requires the end of the space-like singularity, thus inducing the Cauchy horizon.
By definition, one do no calculation beyond the Cauchy horizon. However, we can conjecture that a time-like singularity exists inside the Cauchy horizon.
In this causal structure, a duplication experiment between the Cauchy and the inner horizons is possible \cite{HHSY,HHYZ}.
Free-falling matter can send a signal along the outgoing direction ($A$ in Fig.~\ref{fig1}).
This information can be observed outside the black hole after the information retention time ($B$ in Fig.~\ref{fig1})
thus allowing an observer access to information duplication.
Note that, if we assume a large number of massless degrees of freedom, all processes occur in the semiclassical region.

~\\

{\centering
\section{\label{sec:HM}Horowitz-Maldacena proposal as a selection principle}
}

\begin{figure}
\begin{center}
\includegraphics[scale=0.7]{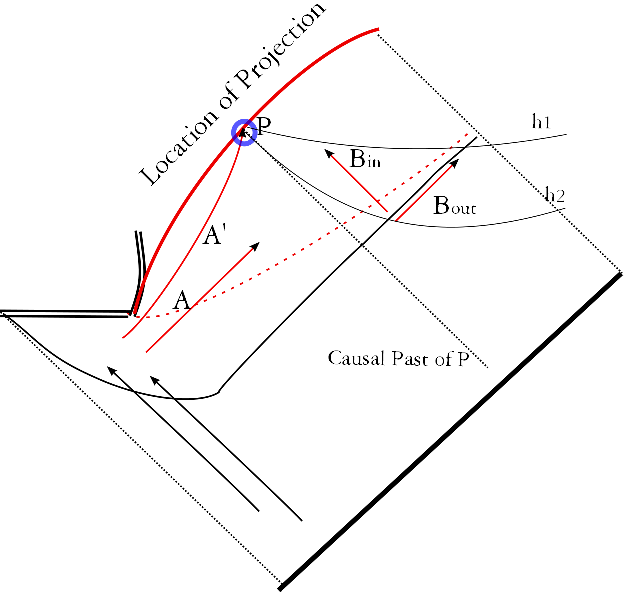}
\caption{\label{fig2}
If one assumes the Horowitz-Maldacena proposal, one needs to specify the location of the projection (red curve).
Information of type $A$ cannot escape to $B_{\mathrm{out}}$, because it is not projected to the projection surface,
but type $A'$ can escape after the collapse at point $P$.
`After' $P$ is not well-defined because it depends on the choice of space-like hypersurfaces
(e.g., $h_{1}$ and $h_{2}$ give different concept of the `after').
However, it cannot be located at the causal past of the point $P$.
Therefore, the duplication experiment becomes impossible through the Horowitz-Maldacena proposal.
Note that, $B_{\mathrm{in}}$ and $B_{\mathrm{out}}$ are maximally entangled.}
\end{center}
\end{figure}

If there is a special relation between the singularity and the outer horizon,
and if the \textit{conspiracy} prevents Hawking radiation ($B$ in Fig.~\ref{fig1}) from containing information related to the out-going information that did not touch a singularity ($A$ in Fig.~\ref{fig1}), black hole complementarity holds.
Although the relation is space-like, if quantum teleportation can be realized in a black hole, a conspiracy can be implemented.
The authors argue that, the Horowitz-Maldacena proposal \cite{HM} can be used exactly for this purpose.

The Horowitz-Maldacena proposal assumes that \cite{HM,GP},
in-falling matter $|i\rangle_{\mathrm{M}}$ becomes entangled with in-going Hawking radiation $|i\rangle_{\mathrm{in}}$ by a unitary transformation $S$ (with $N$ states):
\begin{eqnarray}
|\Psi_{\mathrm{in}}\rangle = \frac{1}{\sqrt{N}} \sum_{i} S|i\rangle_{\mathrm{M}}\otimes|i\rangle_{\mathrm{in}}.
\end{eqnarray}
The in-going Hawking radiation $|j\rangle_{\mathrm{in}}$ is also assumed to be maximally entangled with the outgoing Hawking radiation $|j\rangle_{\mathrm{out}}$:
\begin{eqnarray}
|\Psi_{\mathrm{out}}\rangle = \frac{1}{\sqrt{N}} \sum_{j} |j\rangle_{\mathrm{in}}\otimes|j\rangle_{\mathrm{out}}.
\end{eqnarray}
Now, one strong assumption is that the in-going information is projected because of the final-state assumption near the singularity. Then, we get
\begin{eqnarray}
\langle \Psi_{\mathrm{out}}|\Psi_{\mathrm{in}}\rangle = \frac{S}{N};
\end{eqnarray}
thus unitarity is restored. Of course, the projection near the singularity is not unitary in itself,
which it is similar to the measurement in quantum teleportation.

If the assumption near the singularity is true, each maximally entangled out-going Hawking radiated particle can work as an information carrier,
and after the information retention time \cite{Page,inforretention}, each particle will contain a bit of information.
Proper realization of this mechanism requires the following steps:

\begin{enumerate}
  \item In-going matter collapses to the singularity.
  \item In-going Hawking radiation is projected at the singularity; this is similar to the measurement of quantum teleportation.
  \item By quantum teleportation, each out-going Hawking particle contains information.
\end{enumerate}

Step 3 cannot happen before step 1; hence, we can only see bits of information that have already collapsed the singularity. Then, black hole complementarity is safe.
The global time used to identify the before and the after both inside and outside of a black hole is not well-defined in a general relativistic sense.
It depends on the choice of space-like hypersurfaces (Fig.~\ref{fig2}); this results from the exotic boundary condition of the singularity.
However, we still apply our intuitive logical order in the causal diagram because the future should be located somewhere outside the causal past of the collapsing point ($P$ in Fig.~\ref{fig2}).
Clearly, then, a duplication experiment is not possible, even with the global time ambiguity \cite{HHYZ}.

~\\

{\centering
\section{\label{sec:HP}Thought experiment with the Hayden-Preskill argument}
}

\begin{figure}
\begin{center}
\includegraphics[scale=0.7]{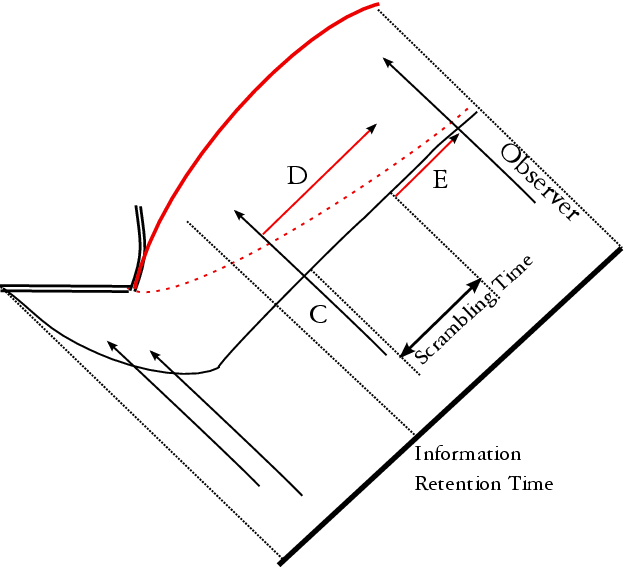}
\caption{\label{fig3}
Degrees of freedom of $C$ are substantially smaller than those of the black hole itself.
Then, according to the Hayden-Preskill argument, information about $C$ must have escaped after the scrambling time.
Then, $D$ and $E$ can be observed by an observer.}
\end{center}
\end{figure}

According to the Hayden-Preskill argument \cite{Hayden:2007cs}, if the black hole \textit{rapidly} mixes or scrambles the states,
whenever some small bits of information fall into a black hole after the information retention time,
the small bits of information almost directly come out, as if being reflected from an information `mirror.'
Thus, as long as the mixing or scrambling time is short enough, one may assume that a duplication experiment is possible.
According to some analysis, the marginal (smallest) rapid mixing time (or, scrambling time) is on the order of
\begin{eqnarray}
t_{\mathrm{scr}} \sim M \log M,
\end{eqnarray}
where $M$ is the black hole mass \cite{Hayden:2007cs,Sekino:2008he}. This time scale can be calculated from the membrane paradigm \cite{Thorne:1986iy,Susskind} and information theoretical arguments \cite{Sekino:2008he}.
If the in-falling information can be sent by a signal along the outgoing direction, and if the outer observer sees that information through Hawking radiation after the scrambling time, falls into the black hole and compares the signals, then eventually, the observation of duplication may be possible. To send a signal with bits of quantum information in the outgoing direction, we need the uncertainty relation $\Delta E \Delta t \sim 1$. If the consistency of black hole complementarity is to be checked in the Kruskal coordinate system, the equation
\begin{eqnarray}
\Delta E \sim \exp{\frac{t_{\mathrm{scr}}}{M}} \geq M
\end{eqnarray}
should hold because the required energy $\Delta E$ is greater than the black hole mass $M$. Of course, for given $t_\mathrm{scr}$, the inequality is marginally valid. In many situations, the scrambling time should be greater than the marginal limit; thus, we can conclude that black hole complementarity is respected in the Hayden-Preskill argument.

Then, what if this argument is applied to a dynamical charged black hole (Fig.~\ref{fig3})?
According to the Hayden-Preskill argument, if one sends small bits of (additional) information ($C$ in Fig.~\ref{fig3}) to the black hole after the information retention time,
one can reconstruct the original information ($C$ in Fig.~\ref{fig3}) from the Hawking radiation ($E$ in Fig.~\ref{fig3}) after the scrambling time.
Let us assume that the in-falling information ($C$ in Fig.~\ref{fig3}) sends a signal along the out-going direction ($D$ in Fig.~\ref{fig3}).
According to the membrane paradigm or the principle of black hole complementarity, it becomes scrambled near the horizon.
That is, it looks like \textit{it does not depend on the inside structure};
thus, according to the Hayden-Preskill argument, it should escape after the scrambling time.
However, one may envisage a situation where, until the scrambling time, $D$ is not projected near the singularity.
Then, the Horowitz-Maldacena selection principle cannot be applied;
the fast scrambling process pushes the information to be escaped to the outside ($E$ in Fig.~\ref{fig3}),
thus invalidating the Horowitz-Maldacena proposal.
The result is that one cannot prevent the observation of duplication in a dynamical charged black hole.
In this sense, a dynamical charged black hole is an unambiguous counterexample of black hole complementarity.

~\\

{\centering
\section{\label{sec:gen}Generalization to the neutral black hole}
}

When we apply and define a duplication experiment and the Horowitz-Maldacena proposal with a neutral black hole, all things are the same (the projection surface will be the space-like singularity), but we need to calculate the time scale of the infalling information $A$ in Fig.~\ref{fig1}. For a charged black hole, the time scale is $\Delta t \sim M$; hence, it is sufficiently long \cite{HHSY}. On the other hand, for a Schwarzschild black hole in four dimensions, the time scale is $\Delta t \sim \exp -M^{2}$, where $M$ is the mass of the black hole. Hence, the time scale is sufficiently small so that the required energy $\Delta E \sim \hbar/\Delta t$ to send a signal is much greater than the black hole mass itself. Therefore, some authors think that such a duplication experiment is not possible \cite{inforretention}.

However, one illustrative example is the $1+1$ dimensional model \cite{Callan:1992rs,deAlwis:1992hv,Bilal:1992kv,Russo:1992ax}. In the authors' previous work \cite{Kim:2013fv}, they calculated the time difference between the event horizon and the apparent horizon along the ingoing null direction by using the RST exact solution \cite{Russo:1992ax}:
\begin{eqnarray}\label{eq:delt}
\Delta t &=& \frac{\kappa}{4} + \frac{M}{\lambda}\left[ 1 - \left( 1 - e^{-4M/\kappa\lambda} \right)^{-1} \right] - \frac{\kappa}{4} \left[ \log \left(1+\frac{\kappa\lambda}{4M} \right) + \log \left(1- e^{-4M/\kappa\lambda} \right) \right]\\
&\simeq& \frac{\kappa}{4},
\end{eqnarray}
where $M$ is the mass of the black hole, $\lambda$ is related to the cosmological constant, $\kappa = N/12$, and $N$ is the number of scalar fields that contribute to Hawking radiation. Therefore, if we introduce sufficient $\kappa$ (note that as $\kappa$ increases, CGHS and RST models become better descriptions), then even between the apparent horizon and the event horizon (hence, outside the event horizon), we have a sufficient time $\Delta t$. Thus, it is also the same for the scrambling time; in general, to do a duplication experiment with the scrambling time, the required time is longer than $\Delta t$ (Eq.~(\ref{eq:delt})). Hence, the thought experiment in Sec.~\ref{sec:HP} is easily allowed, even with a reasonable number of scalar fields. Therefore, we conclude that we can construct a thought experiment that shows the fact that the Horowitz-Maldacena proposal cannot rescue black hole complementarity without assuming an exponentially large number of scalar fields.

~\\

{\centering
\section{\label{sec:dis}Discussions}
}

We discussed the causal structure of a dynamical charged black hole with a large number of massless fields.
If we assume the Horowitz-Maldacena proposal to be true, the black hole complementarity principle seems to be maintained by quantum teleportation.
However, when we introduce a random scrambling of the membrane, the Hayden-Preskill argument causes black hole complementarity to fail.
Furthermore, we can drop the assumption of a large number of scalar fields for two-dimensional neutral black hole models.
Therefore, we conclude that the Horowitz-Maldacena proposal cannot rescue complementarity; hence, it cannot be an alternative to the firewall.

Note that from the very outset, even without the Hayden-Preskill argument,
the Horowitz-Maldacena proposal has some problems.
Interactions between the in-falling matter and the in-going Hawking radiation may violate unitarity \cite{GP}.
Moreover, we still do not know whether, beyond the Cauchy horizon, there is a singularity or a second asymptotic region.
Now, from the Hayden-Preskill argument, it is clear that \textit{the Horowitz-Maldacena proposal cannot solve the information loss problem.}

Hence, if a firewall exists, then it cannot be near the singularity. Rather, the firewall should become a new kind of singularity near the apparent horizon. Therefore, it should be outside the event horizon; hence, the effects of the firewall can be observed by an asymptotic observer \cite{Hwang:2012nn,Kim:2013fv}. The consistency of the firewall or observable effects for an asymptotic observer should be discussed further, and we have left for a future work.

~\\
\section*{Acknowledgment}
The authors would like to thank Ewan Stewart, Dong-il Hwang, and Sungwook Hong for the insightful conversation.
DY and HZ are supported by the BK21 and by the Korea Research Foundation grant (KRF-313-2007-C00164, KRF-341-2007-C00010). DY is supported by the National Research Foundation of Korea grant through the Center for Quantum Spacetime (CQUeST) of Sogang University (No.~2005-0049409) and the JSPS Grant-in-Aid for Scientific Research (A) (No.~21244033). HZ is supported by TUBITAK research fellowship program for foreign citizens.

\end{document}